\documentstyle[epsf]{article}

\def\be{\begin{equation}}
\def\ee{\end{equation}}
\def\bea{\begin{eqnarray}}
\def\eea{\end{eqnarray}}

\begin{document}

\title{{\it AB INITIO} CALCULATIONS OF THE SPIN-HALF {\it XY} MODEL}

\author{D. J. J. FARNELL$^a$ and M. L. RISTIG$^b$}

\date{\today}

\maketitle

\begin{center}

{\it $^a$Department of Physics, University of Manchester Institute of
  Science and Technology (UMIST), P O Box 88, Manchester M60 1QD, United 
  Kingdom}

\vspace{0.25cm}

{\it $^b$Institut f\"ur Theoretische Physik, 
  Universit\"at zu K\"oln, Z\"ulpicher Str., 50937 K\"oln, Germany.}

\end{center}

\begin{abstract}
In this article, the correlated basis-function (CBF) method 
is applied for the first time to the quantum spin-half {\it XY} model 
on the linear chain, the square lattice, and the simple cubic lattice. 
In this treatment of the quantum spin-half {\it XY} model a Jastrow ansatz 
is utilised to approximate the ground-state wave function.
Results for the ground-state energy and the sublattice 
magnetisation are presented, and evidence that the CBF 
detects the quantum phase transition point in this 
model is also presented. 
The CBF results are compared to previous coupled cluster method (CCM) 
results for the spin-half {\it XY}  model, and the two formalisms are 
then compared and contrasted.  
\end{abstract}














In this paper we consider the $T=0$ properties of the quantum spin system 
known as the spin-half {\it XY} model, described by the Hamiltonian
\begin{equation} 
H=\frac{1}{8}\sum_{i,j} \biggl [ 
(1+\gamma)\sigma_i^x \sigma_j^x+(1-\gamma) \sigma_i^y
\sigma_j^y
\biggr ] ~~ ,
\label{eq1} 
\end{equation}
in the regime $0 < \gamma \le 1$. Note that the index $i$ runs 
over all $N$ lattice sites and that the index
$j$ runs over the $z$ nearest-neighbour to $i$ on the linear 
chain ($z=2$), the square lattice ($z=4$), and the cubic 
lattice ($z=6$).

In the regime $0 < \gamma \le 1$ the ground state is believed to exhibit 
N\'eel ordering in the $x$-direction, and for $-1 < \gamma 
\le 0$ the ground state is again believed to possess
N\'eel ordering in the $y$-direction. We note that 
a phase transition point occurs for the linear chain model 
at exactly $\gamma=0$, and that  N\'eel ordering 
is found to disappear at this point. For spatial dimensionality 
greater than one, the phase transition point of the anisotropic 
model is also believed to be at (or very near to) $\gamma=0$ 
from approximate calculations. The ground state of the spin-half 
{\it XY} model on the square and cubic lattices at $\gamma=0$ 
is also believed to be N\'eel-ordered in the $xy$-plane.

The spin-half {\it XY} model was solved exactly by Lieb, Schultz 
and Mattis\cite{Lieb} for the linear chain using the Jordan-Wigner 
transformation. Since then the ground- and excited-state properties 
have been extensively studied by many authors (see, for examples, 
Refs. [2,3]). However, no exact results exist for
higher spatial dimensionality, although approximate results 
such as those from spin-wave theory, \cite{Zheng} Monte-Carlo 
(QMC) methods,\cite{Ding,Zhang} series expansions,\cite{Hamer} 
and the coupled cluster method (CCM) \cite{Farnell} has proven 
to be highly successful. Extrapolated finite size calculations 
\cite{Betts} have also been performed for $\gamma=0$. 


The correlated basis function (CBF) method$^{10-20}$ 
is a widely applied and accurate method of modern-day quantum many-body
theory. Recently, this method has been applied with great success
to the Ising model in a transverse magnetic field at zero 
temperature.$^{21-24}$ 
In this article we wish to apply the CBF method to the spin-half 
{\it XY} model. We begin this process by firstly performing a number of
unitary transformations on the local spin axes on two sublattices
$\{A,B\}$ in order to simplify the problem. The first such transformation 
on the $A$-sublattice is given by,
\begin{equation}
\sigma^x \rightarrow \sigma^z ~~ ; ~~ \sigma^y \rightarrow \sigma^x ~~ ; ~~
\sigma^z \rightarrow \sigma^y ~~ ,
\label{eq2}
\end{equation}
and the second transformation on the $B$-sublattice is given by, 
\begin{equation}
\sigma^x \rightarrow -\sigma^z ~~ ; ~~ \sigma^y \rightarrow -\sigma^x ~~ ; ~~
\sigma^z \rightarrow \sigma^y ~~ .
\label{eq3}
\end{equation}
Note that both of these transformations are simply rotations 
of the local spin-axes of the spins, and that 
the eigenvalue spectrum of the problem is left unchanged because 
these transformations are unitary. The Hamiltonian may now be rewritten in 
terms of these new spin-axes as 
\begin{equation} 
H =  - \frac 18 \sum_{i,j} \biggl [
(1-\gamma) \sigma_{i}^x \sigma_{j}^x +
(1+\gamma) \sigma_{i}^z \sigma_{j}^z 
\biggr ] ~~ .
\label{eq4} 
\end{equation}
We may now define a ground-state trial wave function, given by
\begin{equation}
| \psi \rangle = {\rm exp} \{ U \} ~ |0\rangle ~~ ; ~~ 
U = \frac 12 \sum_{i<j}^N u({\bf r}_{ij}) \sigma_i^x
\sigma_j^x ~~ ,
\label{eq5}
\end{equation}
where $u({\bf r}_{ij})$ is the pseudopotential. The reference state 
$|0\rangle$ is given by a tensor product of spin states
which have eigenvalues of $+1$ with respect to $\sigma^z$,
and this state is an exact ground eigenstate of the Hamiltonian 
Eq. (\ref{eq4}) when $\gamma=1$. Translational invariance also 
implies that the pseudopotential, $u({\bf r}_{ij})$, depends 
only on the relative distance, ${\bf n} = {\bf r}_i - {\bf 
r}_j \equiv {\bf r}_{ij}$.

The treatment of the spin-half {\it XY} model by the CBF 
method is continued by defining the lattice magnetisation 
(i.e., again the magnetisation in the $z$-direction in terms 
of the rotated local spin-axes), 
given by
\begin{equation}
M = \frac {\langle \psi \mid \sigma_i^z \mid \psi \rangle}
{\langle \psi | \psi \rangle} ~~,
\label{eq6}
\end{equation}
for a ground-state trial wave function, $| \psi \rangle$.
Furthermore, the `transverse' magnetisation (in terms 
of the rotated local spin-axes) is given by,
\begin{equation}
A = \frac {\langle \psi \mid \sigma_i^x \mid \psi \rangle}
{\langle \psi | \psi \rangle} ~~.
\label{eq7}
\end{equation}
We may now define a spatial distribution function (which plays 
a crucial part in any CBF calculation) in the following 
manner,
\begin{equation}
G({\bf n}) = \frac {\langle \psi \mid \sigma_i^x \sigma_j^x  \mid \psi \rangle}
{\langle \psi | \psi \rangle} ~~ .
\label{eq8}
\end{equation}
Furthermore, we may also determine an expression for the
expectation value of the ground-state energy of the spin-half 
{\it XY} Hamiltonian of Eq. (\ref{eq4}), where
\begin{equation}
  \frac {E}N = \frac {\langle \psi | H | \psi \rangle}  
  {N \langle \psi | \psi \rangle} ~~ .
\label{eq9}
\end{equation}
In the region $\gamma \ge 0$ we now make the explicit assumption
that $A=0$, which is in agreement with our Ansatz for the trial wave 
function of Eq. (\ref{eq2}). This assumption furthermore implies 
that $A/N \equiv  \langle P^{\rightarrow} \rangle -
\langle P^{\leftarrow} \rangle = \rho^{\rightarrow} -
\rho^{\leftarrow} =0$, where $P^{\rightarrow}$ and $P^{\leftarrow}$
are spin projection operators in the positive and negative
$x$-directions respectively.  However, we note that $\rho^{\rightarrow} 
+ \rho^{\leftarrow} =1$ must also be correct, which therefore 
implies that $\rho^{\rightarrow} = \rho^{\leftarrow} = 1/2$.
Hence, we may treat this problem completely analogously to
a binary-mixture of two types of bosons \cite{Ristig} each with a
density equal to one-half.

\begin{figure}
\epsfxsize=8cm \centerline{\epsffile{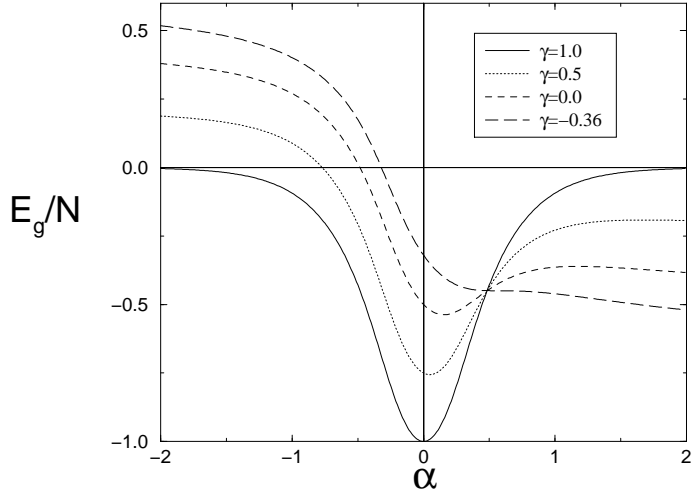}}
\vspace{-0.5cm}
\caption{Results for the ground-state energy of the spin-half {\it XY}
model for the square lattice plotted as a function of the strength of
the nearest-neighbour pseudopotential, $\alpha$, for varying $\gamma$.
At $\gamma=-0.36$ we see that the minimal solution that we have
tracked from $\gamma=0$ is lost.}
\label{fig1}
\end{figure}

The expression in Eq. (\ref{eq9}) may
be determined via a hyper-netted chain (HNC) cluster expansion, and it
is readily found using this procedure that the ground-state energy
is given in terms of a functional with respect to the pseudopotential,
$u({\bf n})$, where
\begin{equation}
\frac EN = -\frac 18 \sum_{\bf n} \Delta({\bf n}) \biggl [ (1-\gamma)
G({\bf n}) + (1+\gamma) ~ M^2 ~ {\rm cosh} [ u({\bf n})] \biggr ]
~~.
\label{eq10}
\end{equation}
Note that $\Delta({\bf n})$ is unity if ${\bf n}$ is a
nearest-neighbour vector and is zero otherwise.

Self-consistent HNC equations may also be determined. These equations
may be then iteratively solved, and thus $G({\bf n})$ (and so the
ground-state energy) may be also obtained.  The first method of finding 
the pseudopotential has a ``variational'' flavour, and we 
parametrise $u({\bf n})$ in the following way,
\begin{equation}
u({\bf n}) = \alpha ~ \Delta({{\bf n}}) ~~.
\label{eq16}
\end{equation} 
$\Delta({\bf n})$ is unity if ${\bf n}$ is a nearest-neighbour
vector and is zero otherwise. We now minimise the ground-state energy with
respect to $\alpha$ at a given value of $\gamma$.  Indeed, at
$\gamma=1$ we already know that all correlations have zero strength as
our reference state $|0\rangle$ is an exact ground eigenstate of
Eq. (\ref{eq4}), and this implies that $\alpha=0$. We thus track this
solution at $\gamma=1$ in the regime $\gamma<1$, and the ground-state
energy as a function of $\alpha$ for various values of $\gamma$ is
plotted in Fig. \ref{fig1} for the square lattice. We may see that at
$\gamma=-0.36$ the minima that we have tracked from $\gamma=1$ become
a point of inflection.

The second such method of determining the pseudopotential is to
determine the optimal value for the function $u({\bf n})$ with respect
to the ground-state energy, $E/N$. This is stated as,
\begin{equation}
\frac {\delta E }{\delta u({\bf n})} = 0 ~~,
\label{eq17}
\end{equation}
which may be determined analytically from Eq. (\ref{eq10}).  In the
context of this article, this approach shall be referred to as the
{\it Paired-Phonon Approximation} (PPA), in analogy with a binary
mixture of two types of bosons, for example.  Note that we do not
explicitly state here the resulting PPA equations for this model,
although the treatment is fully analogous to that performed for the
transverse Ising model and the interested reader is referred to
Ref. [22] 
for a full account of this calculation.

\begin{figure}
\epsfxsize=9cm \centerline{\epsffile{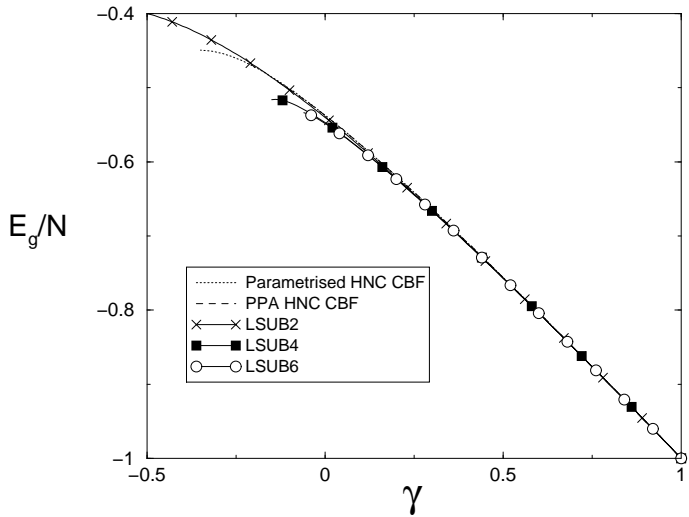}}
\vspace{-0.5cm}
\caption{CBF results for the ground-state energy of the spin-half {\it
XY} model on the square lattice compared to results of high-order CCM
results of Ref. [8].}
\label{fig2}
\end{figure}

For details of the specific application of the CCM to the spin-half
{\it XY} model the interested reader is referred to Ref. [8].
We note however that two types of approximations
are made, namely, the SUB2 approximation which retains all two-body
correlations in the approximate CCM ground-state wave function, and
the LSUB$m$ which retains all correlations in a locale defined by $m$.



Results for the CBF ground-state energy of the spin-half {\it XY} model on
the square lattice compared to results of high-order CCM results are
given in Fig. \ref{fig2} and, for the isotropic point ($\gamma=0$)
only, in Table \ref{tab2}. We may see from Fig. \ref{fig2} that both sets 
of results are in excellent qualitative agreement over a wide range of
$\gamma$. It is furthermore seen from Tables \ref{tab1}-\ref{tab3} that 
CBF results are in excellent quantitative agreement with LSUB2 CCM results 
at $\gamma=0$. This is a perfectly reasonable result because both the 
CBF and CCM LSUB2 results only utilise two-body correlations. 
It is, however, expected that the inclusion of higher-order 
correlations in the CBF trial wave function would produce more accurate
results for the energy, as is seen for the CCM. Thus, from Tables 
\ref{tab1}-\ref{tab3}, 
we see that the CBF results at $\gamma=0$ capture about 59$\%$ of the
correlation energy for the linear chain, 76$\%$ of the correlation
energy for the square lattice and 85$\%$ of the correlation energy for
the cubic lattice (in comparison with exact and extrapolated CCM
results). Indeed, the extrapolated CCM results present some of the most
accurate results yet seen for the isotropic {\it XY} model on the square and 
cubic lattices.  (Results for the linear chain and cubic lattice are 
qualitatively similar to the results presented for the square
lattice in Fig. \ref{fig2} and so are not plotted here.)

\begin{figure}
\epsfxsize=8.5cm \centerline{\epsffile{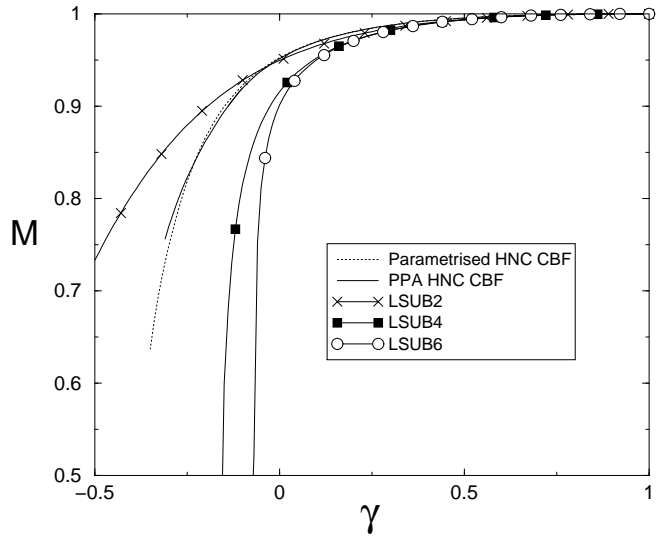}}
\vspace{-0.5cm}
\caption{CBF results for the sublattice magnetisation of the spin-half
{\it XY} model on the square lattice compared to results of high-order
CCM results of Ref. [8].}
\label{fig3}
\end{figure}

Results for the sublattice magnetisation of the spin-half {\it XY}
model on the square lattice are presented in Fig. \ref{fig3} and in
Table \ref{tab2} for the isotropic point, $\gamma=0$.  Again, it is
seen from Fig. \ref{fig3} that the CBF results are in good qualitative
agreement with the known results of this model.  However, the CBF
result for the sublattice magnetisation at the isotropic point
($\gamma=0$) is slightly too high, although it is again expected that 
higher accuracy would be achieved with the inclusion of higher-order
correlations in the approximate CBF ground-state wave
function.  Again, results for the linear chain and cubic lattice are
fully analogous to the square lattice case and so are presented only
for the isotropic model in Tables \ref{tab1} and \ref{tab3}.

\begin{table} 
\caption{Ground-state energy and sublattice magnetisation for the 
one-dimensional {\it XY} model at $\gamma=0$ compared to exact results of
Ref. [3] and CCM results of Ref. [8].  The critical values of $\gamma$
for the anisotropic model are also given.}
\vspace{0.5cm}
\begin{center}
\begin{tabular}{|l|c|c|c|}  \hline\hline 
LSUB{\it n} &E$_g$/N &M &$\gamma_c(n)$                  \\ \hline\hline 
Parametrised CBF &$-$0.29025  &0.8919   &--             \\ \hline 
PPA CBF          &$-$0.29030  &0.8904   &--             \\ \hline 
LSUB2            &$-$0.30381  &0.8373   &--             \\ \hline 
SUB2             &$-$0.31038  &0.7795   &$-$0.10789     \\ \hline 
LSUB$\infty$     &$-$0.31829  &--       &--             \\ \hline 
Exact &$-$0.318310 &0.0 &0 \\ \hline
\end{tabular}
\end{center}
\label{tab1}
\end{table}


Results for the phase transitions points predicted by the CCM
method are also given in Tables \ref{tab1}-\ref{tab3}, although
no such results are explicitly given for the CBF method in these 
tables. It is however noted here that the loss of ``minima'' within 
the parametrized HNC CBF approach (at $\gamma=-0.36$ for the 
square lattice) may be associated with a phase transition
within this system. This constitutes a powerful result for
such a simple variational-style calculation. Note that similar 
behaviour is also seen for both the linear chain and cubic lattices.
An analogous change in the energy surface with respect to $u({\bf n})$
 for the CBF PPA approach seems to occur for varying values of 
$\gamma$. However, in this case, the situation is much less clear-cut 
because, near to this point, convergence of the PPA equations becomes 
very difficult. 


In this article, the CBF method has been applied with much success to
the quantum spin-half {\it XY} model on the linear chain, the square 
lattice, and the cubic lattice in order to obtain accurate results
for the ground-state energy and the sublattice magnetisation. 
These results were found to be in excellent qualitative agreement
with previous CCM calculations,\cite{Farnell} although more quantitatively
accurate CBF results would be possible with the inclusion of
higher-order correlations (than Jastrow correlations) in the trial
ground-state wave function. A strength of the CBF method is that it is
not limited by the presence of frustration, in contrast with QMC
methods for example, and a direct extension of this work would be to
include next-neighbour-neighbour interactions in our model. Indeed,
this presents the possibility that this calculation might be utilised
to provide a trial or guiding wave function for these QMC techniques
in the presence of such frustrating next-nearest-neighbour bonds.
Also, it is possible to see that an extension of this work to
Heisenberg antiferromagnetic (HAF) models could follow a similar path
to that outlined in this article. One would perform a similar set of
rotations of the local spin-axes and then perform HNC re-summations of
the relevant quantities that one is interested in. For example, one
might consider the HAF on the triangular lattice, and in this case one
would perform a rotation on three sublattices -- such as that utilised
by Singh and Huse for this model.\cite{Singh}  Previous CCM results
for the spin-half {\it XY} model quoted in this article were also seen
to provide excellent results for this model, and they are furthermore
a valuable yardstick with which to compare our new CBF results with.

\begin{table} 
\caption{Ground-state energy and sublattice magnetisation for the
square lattice {\it XY} model at $\gamma=0$ compared to CCM
calculations of Ref. [8] and series expansion calculations of
Ref. [7]. The critical values of $\gamma$ for the anisotropic model
are also given, where the value in parentheses is the estimated error
in the final decimal place shown.}
\vspace{0.5cm}
\begin{center}
\begin{tabular}{|l|c|c|c|c|}  \hline\hline 
LSUB{\it n} &E$_g$/N &M &$\gamma_c(n)$                   \\ \hline \hline 
Parametrised CBF &$-$0.53738 &0.9524   &--               \\ \hline 
PPA CBF          &$-$0.53774 &0.9515   &--               \\ \hline 
LSUB2            &$-$0.54031 &0.9496   &--               \\ \hline 
SUB2             &$-$0.54633 &0.9190   &$-$0.030(1)      \\ \hline 
LSUB$\infty$     &$-$0.54892 &0.869    &0.00(1)          \\ \hline 
Series Expansion &$-$0.5488  &0.872    &--               \\ \hline
\end{tabular}
\end{center}
\label{tab2}
\end{table}

We note that the CBF approach utilises a Jastrow
wave function and its bra states are always the explicit Hermitian
adjoint of the corresponding ket state. Hence, for the CBF approach,
an upper bound to the true ground-state energy is, in principle,
obtainable, although the approximations made in calculating the energy
may destroy it. By contrast, the CCM uses a bi-variational approach in
which the bra and ket states are not manifestly constrained to be
Hermitian adjoints and hence an upper bound to the true ground-state
energy is not necessarily obtained. Also, the CCM uses creation
operators with respect to some suitably normalised model state in
order to span the complete set of (here) Ising states. The CBF method,
in essence, uses projection operators to form the Jastrow correlations
with respect to a reference state, $|0\rangle$. In some sense, the CCM
is found to contain {\it less} correlations than the others at
`equivalent' levels of approximation (e.g., the CCM LSUB2
approximation versus Hartree and nearest-neighbour Jastrow
correlations).  
A strength of the CCM is that it is well-suited to the inclusion of  
high-order correlations in the approximate ground-state 
wavefunction (for example, via computational techniques).
Furthermore, the CCM requires no information other than the
approximation in $S$ and $\tilde S$ in order to determine an
approximate ground state of a given system.  The CBF method, however,
may require that only a certain subset of all possible diagrams are
summed over (e.g., the HNC/0 approximation).

\begin{table} 
\caption{Ground-state energy and sublattice magnetisation for the
cubic lattice {\it XY} model at $\gamma=0$ compared to CCM results of
Ref. [8].  The critical values of $\gamma$ for the anisotropic model
are also given, where the value in parentheses is the estimated error
in the final decimal place shown.}
\vspace{0.5cm}
\begin{center}
\begin{tabular}{|l|c|c|c|}                                  \hline \hline 
LSUB{\it n} &E$_g$/N &M &$\gamma_c(n)$                   \\ \hline \hline 
Parametrised CBF    &$-$0.78572  &0.9710  &--            \\ \hline 
PPA CBF             &$-$0.78625  &0.9695  &--            \\ \hline 
LSUB2               &$-$0.78687  &0.9715  &--            \\ \hline 
SUB2                &$-$0.79090  &0.9583  &$-$0.01666(1) \\ \hline 
LSUB$\infty$        &$-$0.79201  &0.948   &0.01(1)       \\ \hline
\end{tabular}
\end{center}
\label{tab3}
\end{table}

We finally note that the results of one method reinforce and sometimes 
elucidate the results of the other, and the application, in parallel, 
of two such methods to the same model can lead to a deeper understanding 
of the behaviour of it.


\end{document}